\documentclass[sigconf,natbib=true,nonacm]{acmart}

\AtBeginDocument{%
  }

\usepackage{graphicx}
\usepackage{booktabs}
\usepackage{tabularx}
\usepackage{array}

\setcopyright{none}
\author{Deo Munduku}
\affiliation{%
  \institution{Paris-Dauphine University, PSL Research University}
  \department{UMR CNRS 7243, LAMSADE}
  \city{Paris}
  \country{France}
}
\email{deo.munduku@lamsade.dauphine.fr}

\author{Elsa Negre}
\affiliation{%
  \institution{Paris-Dauphine University, PSL Research University}
  \department{UMR CNRS 7243, LAMSADE}
  \city{Paris}
  \country{France}
}
\email{elsa.negre@lamsade.dauphine.fr}

\renewcommand{\shortauthors}{Munduku and Negre}

\begin{document}

%%
%% The "title" command has an optional parameter,
%% allowing the author to define a "short title" to be used in page headers.
\title{Learning from How Users Interpret: Personalizing Recommendation Explanation Order}

%%
%% The "author" command and its associated commands are used to define
%% the authors and their affiliations.
%% Of note is the shared affiliation of the first two authors, and the
%% "authornote" and "authornotemark" commands
%% used to denote shared contribution to the research.

%%
%% By default, the full list of authors will be used in the page
%% headers. Often, this list is too long, and will overlap
%% other information printed in the page headers. This command allows
%% the author to define a more concise list
%% of authors' names for this purpose.

%\renewcommand{\shortauthors}{Trovato et al.}

%%
%% The abstract is a short summary of the work to be presented in the
%% article.
\begin{abstract}
Personalized explanations typically adapt their content or presentation
based on preferences, interaction histories, or other user characteristics.
We examine whether regularities inferred from users' prior textual
interpretations can also support personalization of explanation
organization. In a controlled within-participant study with 122
participants, prior interpretations were represented along three
dimensions: information selection, relative salience, and relational
organization, with the relational dimension partially operationalized
through co-mobilization. These representations were aggregated into
fixed historical profiles and used to organize otherwise identical
explanation sentences. Participants encountered four organization procedures: their own
historical profile (OWN), another user's historical profile (DONOR), a
leave-one-participant-out population profile (POP), and a target-aware
language-model procedure using the user's historical profile
(PROFILE-LLM). Post-exposure objective comprehension was the primary outcome.
OWN achieved 86.5\% correct responses, compared with 66.0\% for
DONOR, 63.9\% for POP, and 73.0\% for PROFILE-LLM. Mixed-effects
analyses showed higher comprehension with OWN than with each
alternative procedure. No reliable overall condition effect was detected
for recommendation acceptance. These findings indicate that historical
profiles derived from prior interpretations can provide an actionable
signal for personalizing explanation order while keeping explanatory
content and wording fixed.
\end{abstract}

%%
%% The code below is generated by the tool at http://dl.acm.org/ccs.cfm.
%% Please copy and paste the code instead of the example below.
%%
% AVANT
\ccsdesc[500]{Human-centered computing~User models}
\ccsdesc[500]{Human-centered computing~Empirical studies in HCI}
\ccsdesc[300]{Information systems~Users and interactive retrieval}

%%
%% Keywords. The author(s) should pick words that accurately describe
%% the work being presented. Separate the keywords with commas.
\keywords{personalized explanations, explanation organization,
objective comprehension, historical user profiles,
sentence ordering, human-information interaction}

%\received{20 February 2007}
%\received[revised]{12 March 2009}
%\received[accepted]{5 June 2009}

%%
%% This command processes the author and affiliation and title
%% information and builds the first part of the formatted document.
\maketitle
\section{Introduction}
\label{sec:introduction}

Personalized explanations aim to adapt the information presented to the
characteristics of each user. Existing approaches rely, among other
signals, on preferences, interaction histories, past behaviors, or other
individual characteristics to adapt the content, form, or presentation of
an explanation
\cite{SchneiderHandali2019,NunesJannach2017,KoukiEtAl2019}.
These signals can characterize what a user prefers or values, but they do
not necessarily describe how that user organizes information when
interpreting it. Two users with similar preferences may, for instance,
differ in the elements they mobilize, in the relative importance they
assign to them, or in the pieces of information they consider jointly
\cite{Kintsch1988,johnson_laird_2010}. Previous interpretations constitute a complementary source of information
about the user. When a person expresses an interpretation of a situation,
the resulting response may reveal certain properties of how information is
mobilized: which elements are selected, their relative salience, and which
elements are associated in the expressed interpretation
\cite{CarleyPalmquist1992}. A single interpretation may, however, depend strongly on the particular
situation in which it is produced. When some of these properties recur
across multiple interpretations from the same user, they may constitute
interpretive regularities derived from that user's history. The value of these regularities lies in their potential for reuse. They
may provide a signal complementary to preferences for personalizing the
presentation of new information. In the context of explanations, this
opens a particular possibility: personalizing not only the information
that is presented, but also its organization according to how the user has
previously mobilized information. Prior work on personalized explanations has examined adaptations of
content, form, and presentation
\cite{SchneiderHandali2019,NunesJannach2017,KoukiEtAl2019}. Other approaches also exploit user history to adapt generation or model
behavior
\cite{SalemiEtAl2024LaMP,TanEtAl2024OPPU,
LiZhangChen2023PersonalizedPromptLearning}.
More directly, work on information ordering has shown that explanation
organization can itself serve as a dimension of personalization when
guided by priorities explicitly stated by the user
\cite{MundukuNegre2026StructureMatters}. 

To our knowledge, interpretive regularities derived from prior textual
interpretations have not yet been exploited as a signal for personalizing
explanation organization. We therefore ask the following central research
question:
\begin{quote}
\textbf{RQ.} Does organizing an explanation from interpretive regularities
derived from the user's prior interpretations improve objective
comprehension of the presented information compared with alternative
organization procedures?
\end{quote}

To investigate this question, we represent prior interpretations along
three dimensions: information selection, relative salience, and
relational organization, with the relational dimension partially
operationalized through co-mobilization. These dimensions allow us to
construct, for each user, a historical profile reflecting regularities
observed across their prior interpretations and to use this profile to
personalize the organization of new explanations. We evaluate this
approach in a controlled study in which objective comprehension is the
primary outcome, considered separately from recommendation acceptance
and subjective judgments of the explanation
\cite{TintarevMasthoff2012Effectiveness,NimmoEtAl2024}. Our contributions
are threefold:

\begin{enumerate}

    \item We show, in a controlled study with 122 participants, that
organizing an explanation from interpretive regularities derived
from the user's prior interpretations improves objective
comprehension compared with three alternative organization
procedures.

    \item We use a fixed historical profile as an actionable signal
for ordering explanation sentences while holding explanation
content and wording constant, and compare the resulting procedure
with another user's historical profile, a leave-one-participant-out
population profile, and a target-aware LLM reference procedure.

   \item We evaluate recommendation acceptance separately and detect
no reliable overall condition effect in the present study, thereby
distinguishing the observed comprehension advantage from the
acceptance outcome.
\end{enumerate}

\section{Related Work}
\label{sec:related-work}

\subsection{Personalized Explanations: Content, Form, and Organization}
\label{sec:rw-personalized-explanations}

Research on personalized explanations distinguishes several
targets of adaptation. Schneider and
Handali~\cite{SchneiderHandali2019} identify complexity,
decision information, and presentation, and distinguish these
targets from the user information used to personalize them.
This separates what a system knows about a user from
the explanation properties that this knowledge may guide.
In recommender systems, personalization can concern the
arguments used to explain an item's relevance.
Kouki et al.~\cite{KoukiEtAl2019} generate personalized
explanations from multiple sources, including user
interactions and social connections, and compare explanation
styles, combinations of styles, and textual or visual formats. Earlier work proposed considering users' interpretive schemas
as a basis for contextualizing recommendation explanations
\cite{Munduku2025Contextualisation,
MundukuNegre2026Contextualisation}. This conceptual perspective motivates attention to how users
interpret explanatory information.
Munduku and Negre~\cite{MundukuNegre2026Bridging}
also examine user interpretive schemas in recommender
system explanations.
Their work on time-constrained explanations
\cite{MundukuNegre2026StructureMatters}
isolates organization more directly by varying information
order while holding explanation content and wording constant.
Personalized ordering follows the importance explicitly
assigned to information by the user, and its effects
depend on the time available to consult the explanation. These contributions motivate attention to interpretation
and information organization in personalized explanations.
However, the dimension being adapted must be distinguished
from the evidence used to adapt it.
Explicitly stated priorities and regularities derived from
earlier textual interpretations provide different kinds of
information about the user. Evidence supporting organization
based on the former does not establish the usefulness of
organization based on the latter. Our study examines whether
historical interpretations provide a useful basis for
personalizing sentence order in a subsequent explanation.

\subsection{From User History to Explanation Adaptation}
\label{sec:rw-history-adaptation}

Historical data can support personalization through different
mechanisms. LaMP retrieves relevant entries from user profiles
to condition classification and generation
tasks~\cite{SalemiEtAl2024LaMP}, whereas OPPU combines
user-specific parameter-efficient adaptation with retrieval
and textual profiles~\cite{TanEtAl2024OPPU}. For recommendation
explanations, PEPLER uses personalized prompt learning associated
with users and items
\cite{LiZhangChen2023PersonalizedPromptLearning}.
Logic-Scaffolding selects relevant historical items, extracts
their aspects, and generates explanations through guided
reasoning~\cite{RahdariEtAl2024LogicScaffolding}.
ReasoningRec constructs profiles of preferences and dislikes
and uses synthetic explanations to train a model to recommend
and explain~\cite{BismayEtAl2025ReasoningRec}. Other user
properties can also guide adaptation: in human--robot
interaction, Gebell\'{\i} et al.~\cite{GebelliEtAl2025}
use knowledge--memory models to adjust explanation detail
and refer to previously encountered concepts. The relevant distinction is therefore not between explicit
profiles and LLM-based approaches, but between what a user
representation captures and how it guides an explanation.
The studies above rely on properties such as preferences,
item aspects, and acquired knowledge. We instead examine
information selection, relative salience, and relational
organization in historical textual interpretations as inputs
to an ordering procedure, using co-mobilization as a partial
empirical operationalization of the relational dimension. We additionally compare this rule-based procedure with an
LLM that receives the same structured historical profile,
together with target-specific context, and reorders the same
predefined explanation sentences.

\subsection{Evaluating Personalized Explanations:
Comprehension and Decisions}
\label{sec:rw-evaluation}

Explanation evaluations address objectives that are not
interchangeable. Tintarev and
Masthoff~\cite{TintarevMasthoff2012Effectiveness} distinguish,
among others, effectiveness, satisfaction, transparency,
persuasiveness, and efficiency. Their studies show that
personalizing feature-based explanations can increase
satisfaction while reducing measured decision effectiveness.
Alignment with a user profile therefore does not guarantee
benefits across these objectives. Presentation choices also involve trade-offs.
Gedikli et al.~\cite{GedikliJannachGe2014IJHCS} find that
content-based tag-cloud explanations can improve perceived
transparency and satisfaction while requiring greater
cognitive effort. Nimmo et al.~\cite{NimmoEtAl2024}
examine actual understanding of a system providing
explanations and find that few of the individual
characteristics studied are associated with this outcome
in their experimental setting. These findings motivate
direct assessment of whether the properties selected for
personalization provide the intended benefit to users. Generated-explanation studies provide complementary evidence:
PEPLER evaluates textual quality and feature coverage
\cite{LiZhangChen2023PersonalizedPromptLearning}, while
Logic-Scaffolding includes human judgments of relevance,
readability, and factuality
\cite{RahdariEtAl2024LogicScaffolding}. Such measures
characterize explanation outputs but do not directly establish
their effects on users' comprehension. We consequently evaluate
objective comprehension separately from effects on decisions.
A change in choice or recommendation acceptance is not, by
itself, evidence that an explanation has been better understood.

\paragraph{Positioning.}

The Interpretive Schema Manifestation Approach (ISMA)
\cite{AnonymousISMA2027} characterizes expressed interpretations
along three dimensions: information selection, relative salience,
and relational organization. We use this representation to
construct a fixed historical profile from prior interpretations. The present study does not introduce this framework, nor does it
re-evaluate the existence, person-specificity, or cross-situational
informativeness of these interpretive regularities. It instead
addresses a distinct downstream question: whether a historical
profile based on this representation can serve as an actionable
signal for organizing explanatory information. Specifically, we use the user's historical profile to determine
the order of otherwise identical explanation sentences and test
whether this organization improves objective comprehension
relative to another user's historical profile, a
leave-one-participant-out population profile, and a target-aware
LLM procedure. Recommendation acceptance is examined separately.

\section{History-Derived Explanation Organization}
\label{sec:proposed-method}

We use ISMA-derived representations as input to a
history-based explanation-organization procedure. Historical
interpretations are first represented and aggregated into a fixed
historical profile before explanation evaluation. The procedure
examined in the present study then maps this historical profile to
the sentence order of a new explanation. For each target situation,
the explanation consists of six predefined sentences. Their content,
wording, and typography remain unchanged: only their order varies.

\subsection{Historical Trace Representation}
\label{sec:structural-representation}

Historical traces are represented along the three ISMA dimensions
of expressed interpretation: information selection, relative salience,
and relational organization. Selection concerns which information
is mobilized in a response; salience characterizes its relative
prominence within that response; relational organization concerns
how information units are associated. Each historical trace is a written response produced in a source
situation presenting six information units. These units are aligned
across situations through six functional roles: cost or price,
access or travel, core attributes of the offering, experience
conditions, availability, and usage or contractual conditions. Core attributes describe what the offering includes, provides,
or enables, such as an activity's content or an event's
programme. The role labels are adapted to the materials of the
present study and are intended to preserve functional alignment
rather than exact lexical identity with earlier operationalizations.
This functional alignment provides common coordinates without
assuming semantic equivalence across domains. Throughout this
section, \(u\) denotes a user and \(t\) a source episode. The indices
\(i,j\in\{1,\ldots,6\}\), with \(i\neq j\), identify distinct
functional roles and their corresponding information units within
an episode.
A hat denotes a quantity estimated from textual responses. For each historical trace, the extraction procedure produces an
empirical representation
\[
\widehat{\Gamma}_{u,t}
=
\left(
\widehat{\alpha}_{u,t},
\widehat{\omega}_{u,t},
\widehat{\rho}_{u,t}
\right).
\]

These indicators characterize properties of expressed
interpretations; they neither directly observe a latent
interpretive schema nor fully reconstruct the meaning
attributed by the user. Responses undergo Unicode NFKC normalization, whitespace
normalization, and segmentation at sentence-ending
punctuation and line breaks. Segments containing fewer
than four words are discarded. Retained segments and
presented information units are encoded using
\texttt{paraphrase-multilingual-MiniLM-L12-v2}
within Sentence Transformers
\cite{reimers2019sentencebert,
ReimersGurevych2020Multilingual}.

\paragraph{Selection.}

An information unit is identified as mobilized when its cosine
similarity with at least one retained segment reaches the fixed
threshold of \(0.46\). This threshold was determined before the
present study in a separate development calibration
\cite{AnonymousISMA2027} against human annotations,
using cross-validation grouped by participant. The resulting
configuration is applied here without recalibration on the present
dataset. Failure to detect an information unit does not establish
that the user did not consider it.

\paragraph{Relative salience.}

For each selected unit, we retain its maximum similarity
to the retained segments. These values are divided by their
sum over selected units; unselected units receive zero.
This indicator is a proxy for textual salience, rather than
explicitly declared importance, and depends on the
selection procedure.

\paragraph{Co-mobilization.}

We use Qwen3-8B~\cite{yang2025qwen3} to identify the
information units mobilized in each retained segment.
The model receives the segment and the six reference
information units and returns the identifiers of the
units detected. For two selected units in a historical trace
containing retained segments, the empirical co-mobilization
coordinate is the proportion of segments in which both units
are detected:

\begin{equation}
\widehat{\rho}_{u,t,ij}
=
\frac{
\left|
\left\{
s \in S_{u,t} : i,j \in U(s)
\right\}
\right|
}{
|S_{u,t}|
}.
\label{eq:co-mobilization}
\end{equation}

Here, \(S_{u,t}\) contains the retained segment instances,
and \(U(s)\) contains the information-unit identifiers
detected in segment \(s\). The numerator counts segments
in which both units are detected; the denominator counts
all retained segments. This expression applies when both units are selected and
\(|S_{u,t}|>0\). If at least one of the two units is unselected,
the corresponding relational coordinate is left undefined.
If the coordinate is defined and equals zero, both units were
selected, but they were never jointly detected in any retained
segment. The representation contains six selection coordinates,
six salience coordinates, and fifteen possible relational
coordinates, some of which may be undefined. The co-mobilization indicator is symmetric and untyped: it captures
joint detection without specifying relation direction or semantic
type, such as causality, opposition, or compensation.
It therefore provides a partial empirical operationalization
of relational organization. The extraction configuration was fixed before explanation
evaluation and is not adjusted using comprehension, acceptance,
or other evaluation outcomes. The present study evaluates the
explanation-organization procedure constructed from the resulting
historical profiles; it does not provide a new human-annotation
validation of extraction accuracy on this corpus. Model versions,
prompts, calibration details, extraction settings, and the handling
of edge cases, including responses without retained segments and
invalid model outputs, are reported in the supplementary material.

\subsection{Historical Profile Construction}
\label{sec:historical-profile}

For each user, the four source episodes yield four
episode-level representations. These representations
are aggregated into a single historical profile,
comprising selection, relative salience, and relational
components, denoted by
\(\widehat{h}^{\alpha}_{u}\),
\(\widehat{h}^{\omega}_{u}\), and
\(\widehat{h}^{\rho}_{u}\), respectively.

\paragraph{Historical selection and salience.}

For selection and salience, each episode-level vector
is first normalized by its Euclidean norm. The four
normalized vectors are then averaged, and the
resulting vector is normalized again:

\begin{equation}
\widehat{h}^{z}_{u}
=
\mathcal{N}_{2}\left(
\frac{1}{4}
\sum_{t \in T^{\mathrm{hist}}_{u}}
\mathcal{N}_{2}(\widehat{z}_{u,t})
\right),
\qquad z \in \{\alpha,\omega\}.
\label{eq:historical-selection-salience}
\end{equation}

Here, \(T^{\mathrm{hist}}_{u}\) contains the four
source-episode indices for user \(u\), and
\(\widehat{z}_{u,t}\) denotes the corresponding
six-dimensional selection or salience vector.
Normalization is defined as

\begin{equation}
\mathcal{N}_{2}(v)
=
\begin{cases}
v/\lVert v\rVert_{2}, & \lVert v\rVert_{2}>0,\\
\mathbf{0}, & \lVert v\rVert_{2}=0.
\end{cases}
\label{eq:profile-normalization}
\end{equation}

Zero vectors are retained in the four-episode
average. These historical components summarize the
relative structure of the extracted information;
their coordinates are not probabilities or
explicitly reported importance ratings.

\paragraph{Historical co-mobilization.}

Each empirical relational coordinate is averaged over the
source episodes in which it is defined:

\begin{equation}
\widehat{h}^{\rho}_{u,ij}
=
\frac{1}{|T^{\mathrm{def}}_{u,ij}|}
\sum_{t \in T^{\mathrm{def}}_{u,ij}}
\widehat{\rho}_{u,t,ij},
\qquad |T^{\mathrm{def}}_{u,ij}|>0.
\label{eq:historical-co-mobilization}
\end{equation}

The set \(T^{\mathrm{def}}_{u,ij}\) contains the
source episodes in which the relational coordinate for
roles \(i\) and \(j\) is defined. Each contributing
episode has equal weight. Defined zero values contribute
to the average, whereas undefined values are omitted.
A coordinate that is undefined in all four source episodes
remains undefined in the historical profile. No additional
normalization is applied to the relational component. For each relational coordinate, we also record
\(m_{u,ij}=|T^{\mathrm{def}}_{u,ij}|\), the number
of episodes contributing a defined value, and
\(m^{+}_{u,ij}\), the number of these episodes
with positive co-mobilization. These counts
document historical support without changing
the coordinate's value or its eligibility for
the ordering procedure. A positive historical
coordinate may therefore be supported by only
one episode. The resulting historical profile summarizes regularities
in the four observed interpretations; it is not assumed
to represent an immutable psychological trait. Historical
profiles are computed and fixed before explanation
construction. Outcomes from the subsequent evaluation
are not used to construct or update them.

\subsection{From Historical Profiles to Explanation Order}
\label{sec:explanation-order}

Each sentence in the target explanation is associated with one
of the six functional roles. The adaptation rule uses this
correspondence to map the historical profile to a sentence
order. All sentences are retained and appear exactly once,
without added boldface, color, or boxes.

\paragraph{Initial ordering.}

Sentences are sorted by decreasing historical salience.
Ties are broken by decreasing historical selection, followed
by increasing role identifier, fixed in advance.
Selection therefore acts as a secondary criterion; it does not
have a role equivalent to salience in determining the order.

\paragraph{Pair selection.}

Among pairs with defined and strictly positive historical
co-mobilization, we select the pair with the highest value.
For each pair, the first member is the sentence appearing earlier
in the initial order, and the second member is the sentence
appearing later. Tied pairs are ranked by the initial position
of their first member, with earlier positions preferred.
If these positions are identical, the initial positions of
their second members are compared in the same way.

\paragraph{Final ordering.}

The second member of the selected pair is placed immediately
after the first. The relative order of the two members and
that of the remaining four sentences are preserved.
If the pair is already consecutive, no displacement occurs.
If no eligible pair is available, the initial order is retained.
Table~\ref{tab:ordering-example} illustrates this transformation.
The labels denote sentence roles; the sentences themselves
are not rewritten.

\begin{table}[t]
\centering
\small
\caption{Illustrative transformation from the initial order
to the final explanation order.}
\label{tab:ordering-example}
\begin{tabular}{@{}p{0.22\linewidth}p{0.72\linewidth}@{}}
\hline
Stage & Sentence order \\
\hline
Initial order &
Price \(\rightarrow\) Core attributes
\(\rightarrow\) Experience conditions
\(\rightarrow\) Availability
\(\rightarrow\) Usage conditions
\(\rightarrow\) Access \\[3pt]

Selected pair &
Price--Access \\[3pt]

Final order &
Price \(\rightarrow\) Access
\(\rightarrow\) Core attributes
\(\rightarrow\) Experience conditions
\(\rightarrow\) Availability
\(\rightarrow\) Usage conditions \\
\hline
\end{tabular}
\end{table}

In this example, access information moves from the sixth to the
second position. Bringing the pair together therefore gives
local priority to proximity between two information units,
at the expense of modifying the order based on their individual
salience. This choice constitutes a design hypothesis: presenting two
historically co-mobilized information units consecutively may
facilitate their joint consideration. It does not follow
necessarily from ISMA, and its benefit remains to be evaluated.
In particular, the selected pair may rely on a single positive
observation; placing its members consecutively establishes
neither a repeated association nor a particular semantic
relation. For a given historical profile and the same six functional roles,
the rule produces the same role order across target situations.
It thus transfers an organization derived from historical traces
without reassessing the semantic requirements of each target
situation. The evaluation presented in Section~4 concerns this
complete explanation-organization procedure and its effects on
objective comprehension and recommendation acceptance. It does
not, by itself, attribute any observed benefit to individual
historical-profile components.

\section{Experimental Study}
\label{sec:experimental-study}

This study investigates whether a historical profile derived from
users' prior interpretations can guide explanation organization in
new situations. The evaluation focuses on post-exposure objective
comprehension and stated recommendation acceptance while holding
explanation content constant and varying only sentence order.
We distinguish \emph{source situations}, used to collect written
interpretations and construct historical profiles, from
\emph{target situations}, reserved for the evaluation of explanations
organized from those profiles. Target situations were distinct from
those used during profile construction.

\subsection{Participants and Profile Construction}
\label{subsec:participants-profiles}

Profile-construction data were collected specifically for the
present study between August 30 and September 21, 2026. A total
of 135 participants began this phase. Of these, 122 provided the
four written interpretations required to construct a historical
profile, yielding 488 source episodes. Among these participants,
80 completed the tasks in French and 42 in English.
65\% identified as men, 30\% as women, and
5\% as non-binary. Participants were recruited in university lecture halls,
through social media, and within a doctoral-student discussion
group. Participation was voluntary, unpaid, and unrelated to
course assessment or academic grades. All participants were
at least 18 years old and provided informed consent before
taking part. No formal ethics-committee approval was obtained. Participants were identified only through pseudonymous study
codes, and no directly identifying personal information was
collected. Participants were informed about the purpose and
procedure of the study before providing their responses.
Individual written responses were not published, and the study
codes were used only to link responses across the different
phases of the study. The participant sample and data collection were specific to
the present study and did not overlap with the participant
sample used in the separate study underlying the ISMA
reference~\cite{AnonymousISMA2027}. Each participant completed four interpretation tasks in a
counterbalanced order. Within each source situation, the
presentation order of the six information units was also
counterbalanced across participants. The situations concerned
an outdoor activity, a meal-kit delivery service, a mobile-phone
plan, and a home-cleaning service. For each task, participants
read a context accompanied by six information units and then
wrote an open-ended interpretation.

%Figure~\ref{fig:interpretation-interface} illustrates the
%interface used to collect these written interpretations.

%\begin{figure*}[t]
%    \centering
%    \includegraphics[width=0.90\textwidth]{interpretation.png}
%    \caption{Interface for collecting written interpretations,
%    showing the context, six information units, and response field.}
%    \label{fig:interpretation-interface}
%    \Description{An interpretation-task interface containing
%    a situation context, six information units, and a text field
%    for an open-ended written response.}
%\end{figure*}

Participants were not asked to select information units,
assign importance scores, or explicitly identify relationships
between units. The resulting responses therefore constitute
historical textual traces of expressed interpretation rather
than direct reports of the structural dimensions later inferred
from them. Each information unit corresponded to one of the six functional
roles defined in Section~\ref{sec:structural-representation}.
This alignment provides common coordinates across situations
without assuming semantic equivalence across domains.
Information selection and relative salience were estimated using
the fixed extraction procedure described in
Section~\ref{sec:structural-representation}; relational organization
was partially operationalized through the extracted co-mobilization
indicator. Processing the 488 responses involved 2,153 textual
segments, with no parsing failures in the extraction outputs. The four source responses of each participant were aggregated
according to Section~\ref{sec:historical-profile}, yielding
122 historical profiles. All profiles had nonzero selection
and salience components. Historical profiles were computed and
fixed before explanation construction. Evaluation outcomes,
including comprehension responses, subjective judgments, and
recommendation-acceptance responses, were not used to construct
or update them. All 122 participants subsequently completed the
explanation-evaluation phase between September 23 and
September 30, 2026. The same study code linked each evaluation
session to the corresponding historical profile.

\subsection{Explanation Construction and Organization Conditions}
\label{subsec:materials-conditions}

\paragraph{Target situations and explanation content.}

Four target situations involving leisure and cultural activities
were used for evaluation. They were distinct from the source
situations used for profile construction and are summarized in
Table~\ref{tab:target-situations}.

\begin{table}[t]
    \centering
    \small
    \setlength{\tabcolsep}{3pt}
    \renewcommand{\arraystretch}{1.12}
    \caption{Target situations used for explanation evaluation.}
    \label{tab:target-situations}
    \begin{tabularx}{\columnwidth}{
        @{}l
        >{\raggedright\arraybackslash}X
        >{\raggedright\arraybackslash}X@{}
    }
        \toprule
        ID & Decision context & Recommendation \\
        \midrule
        T1 &
        Watch a film in a new city &
        Screening of \emph{Horizon 47} \\
        T2 &
        Choose a cultural activity &
        \emph{Cartographies de lumi\`ere} exhibition \\
        T3 &
        Participate in a cooking workshop &
        \emph{Saveurs du Levant} \\
        T4 &
        Attend a musical performance &
        Ensemble Nova concert \\
        \bottomrule
    \end{tabularx}
\end{table}

Each target explanation contained six predefined sentences,
one per functional role. Historical-profile coordinates were
mapped to target sentences through these roles, relying on
functional alignment rather than semantic identity across
domains. Explanations were available in French and English and
were presented in the language used by the participant during
profile construction. For a given participant--target-situation combination, all four
conditions used exactly the same six sentences. No information
was added, removed, or reformulated; only sentence order varied.
Typography, layout, and visual presentation were otherwise
identical. Sentence lengths ranged from 9 to 16 words in French
(median 13) and from 7 to 17 words in English (median 12).
Detailed sentence lengths and the full bilingual materials
will be provided in the supplementary material.

\paragraph{Organization conditions.}

The four conditions differed in the profile used and in the
procedure determining sentence order, as summarized in
Table~\ref{tab:organization-conditions}.

\begin{table}[t]
    \centering
    \small
    \setlength{\tabcolsep}{3pt}
    \renewcommand{\arraystretch}{1.12}
    \caption{Explanation-organization conditions.}
    \label{tab:organization-conditions}
    \begin{tabularx}{\columnwidth}{
        @{}l
        >{\raggedright\arraybackslash}X
        >{\raggedright\arraybackslash}X@{}
    }
        \toprule
        Condition & Profile source & Ordering procedure \\
        \midrule
        OWN &
        User &
        Fixed rule from
        Section~\ref{sec:explanation-order} \\
        DONOR &
        Another participant, same language &
        Same rule as OWN \\
        POP &
        Other 121 participants, aggregated &
        Same rule as OWN \\
        PROFILE-LLM &
        User &
        Mistral with target context,
        recommendation name, and six sentences \\
        \bottomrule
    \end{tabularx}
\end{table}

\paragraph{OWN: user's historical profile.}

The deterministic ordering rule defined in
Section~\ref{sec:explanation-order} was applied to the
user's historical profile. The resulting role order was
then applied to the four target situations for that participant.

\paragraph{DONOR: another participant's historical profile.}

The same ordering procedure was applied using the historical
profile of another participant from the same language group.
Assignment was bijective and excluded self-assignment, so that
each profile served exactly once as a donor within its language
group. Donor assignments were fixed across target situations
and before evaluation outcomes were collected. Full assignment
details and the random seed will be documented in the
supplementary material.

\paragraph{POP: population profile.}

For each user, a leave-one-participant-out population
profile was constructed from the other 121 participants.
The historical-profile selection and salience components of the
other 121 participants were averaged and L2-normalized; relational
coordinates were averaged over participants for whom they were
defined, retaining defined zeros and excluding undefined values. The OWN ordering rule was
then applied to this population profile. POP therefore provided a population-level historical reference,
whereas DONOR used another participant's historical profile.

\paragraph{PROFILE-LLM: target-aware ordering from the user's historical profile.}

Mistral-7B-Instruct-v0.3~\cite{jiang2023mistral} received the
same structured historical profile used in OWN, together with
the target context, recommendation name, and six predefined
sentences. The prompt distinguished undefined relational
coordinates from coordinates defined with a zero value. The model did not generate or rewrite sentence content; it
returned only a permutation of identifiers 1--6, from which
the explanation was assembled. All 488 PROFILE-LLM outputs
were valid permutations. The complete prompt, model revision,
and generation parameters will be provided in the
supplementary material. Unlike OWN, DONOR, and POP, which produced a fixed role order
for a given historical profile, PROFILE-LLM could produce
target-specific orders because it additionally received target
context and sentence content. The comparison therefore concerns
complete explanation-organization procedures rather than an
isolated effect of the language model.

\paragraph{Formative pretest.}

Twelve pretest participants, distinct from the 122 main-study
participants, completed a formative pretest, with six participants
per language. Each examined four explanations and rated clarity,
coherence, and naturalness on three-point scales while also
providing free-text comments. Mean ratings were 3.00 for
clarity and 2.21 for both coherence and naturalness in each
language. The pretest informed clarification of the third functional-role
label and correction of one English sentence, after which
PROFILE-LLM orders were regenerated using the finalized
materials. Because the pretest concerned an earlier material
version, it was treated as formative rather than as validation
of the final explanation sequences. Detailed pretest materials
will be provided in the supplementary material.

\paragraph{Checks on constructed explanations.}

A total of 1,952 candidate explanations were constructed
(122 participants $\times$ 4 target situations $\times$
4 conditions), of which 488 were actually presented during
evaluation. All candidates were valid permutations and preserved
identical sentence content across conditions for a given
participant--target-situation combination. The four candidate orders were distinct in 478 of the 488
combinations; the remaining ten comprised five
PROFILE-LLM--OWN and five PROFILE-LLM--DONOR matches, all
retained. Across OWN profiles, the selection tie-breaker changed none
of the 122 initial orders. The relational grouping step
modified 28 of the 122 orders; in the remaining 94, the
selected pair was already adjacent. Thus, historical salience
accounted for most observed sentence-order structure, while
the relational step modified a subset of cases. Detailed
construction diagnostics will be provided in the supplementary
material.

\subsection{Experimental Design and Procedure}
\label{subsec:design-procedure}

All 122 participants in the profile-construction sample
completed the explanation-evaluation session between
September 23 and September 30, 2026. The evaluation followed
a within-participant design with four conditions: OWN, DONOR,
POP, and PROFILE-LLM. Each participant encountered each target
situation and each organization condition exactly once, yielding
four trials per participant and one explanation version per
target situation.

\paragraph{Assignment and counterbalancing.}

Four condition--situation assignments defined by a Latin
square were crossed with four cyclic rotations of
T1--T2--T3--T4, producing 16 experimental schedules.
This distributed conditions across target situations and
trial positions, although it did not balance every possible
carryover sequence. Because each target situation used the
same comprehension questions across organization conditions,
this counterbalancing also distributed the fixed comprehension
questions across conditions. Participant codes were randomly assigned to schedules within
language groups subject to predefined allocation counts.
For the 80 French-speaking participants, each schedule
received five participants; among the 42 English-speaking
participants, ten schedules received three participants and
six received two. The allocation and randomization were fixed
before evaluation responses were collected. Participants were
not shown condition labels. Complete assignment schedules and
randomization details will be documented in the supplementary
material.

\paragraph{Procedure.}

Source-data collection, historical-profile construction,
donor assignment, population-profile construction, and
candidate-explanation generation were completed before
evaluation responses were collected. Evaluation outcomes
were not used to modify these objects. The explanation-evaluation phase took place between
September 23 and September 30, 2026, in a separate session
using the same study code to link responses to the corresponding
historical profile. For each trial, participants read the
target context, recommendation, and explanation in the same
language as their source interpretation tasks. Reading was self-paced, with no imposed time limit.
Explanation-display duration was automatically recorded from
explanation onset to the click that opened the evaluation
questions; it therefore measures exposure duration rather
than necessarily active reading time. After the questions were opened, the explanation was removed
and could not be revisited. Participants first reported
recommendation acceptance, then answered the factual and
integration comprehension questions, and finally completed
the three subjective judgments described in
Section~\ref{subsec:measures}. The procedure was repeated
across the four trials, with no correctness feedback provided
during the session.

\subsection{Measures}
\label{subsec:measures}

Each trial collected six responses: one recommendation-acceptance
response, two objective-comprehension responses, and three
subjective judgments, yielding 24 questionnaire responses per
participant across four trials. Explanation-display duration
was recorded automatically and was not an additional
questionnaire item.

\paragraph{Recommendation acceptance.}

Immediately after the explanation was removed, participants
answered: ``In the situation presented, would you accept this
recommendation?'' Response options were ``No,'' ``Uncertain,''
and ``Yes,'' with uncertainty retained as a separate category. Acceptance was collected before the comprehension questions so
that those questions would not first direct attention toward
particular information. It represents stated acceptance within
the experimental scenario and is not, by itself, interpreted
as evidence of a better decision.

\paragraph{Post-exposure objective comprehension.}

Each target situation included two multiple-choice questions.
The first assessed a factual detail relevant to the
recommendation; the second required combining at least two
pieces of information to identify a practical consequence or
applicable condition. Each question had one correct answer and
plausible distractors. For a given situation and language, questions and response
options were identical across organization conditions.
Response options were randomly permuted except when a meaningful
numerical or temporal progression required a fixed order.
Correct responses were coded as 1 and incorrect responses as 0;
missing responses remained distinct from incorrect responses. A descriptive 0--2 trial score was also computed when both
questions were answered. The primary objective outcome was
question-level correctness across the eight objective-comprehension
questions. Differences between factual and integration questions
were examined in the secondary analysis described in
Section~\ref{subsec:analysis-plan}. Because participants responded after the explanation had been
removed, performance reflects both comprehension and short-term
retention and does not identify a unique cognitive mechanism. The eight questions were piloted with the same twelve formative
pretest participants, who were distinct from the main sample.
The pilot assessed clarity, uniqueness of the correct answer,
and plausibility of distractors. Because these participants had
already encountered the explanations, it was not used to
estimate first-exposure question difficulty. Final questions,
response options, and answer keys were fixed before the main
evaluation.

\paragraph{Subjective judgments.}

After the objective questions, participants responded in a
fixed order to three items:

\begin{enumerate}
    \item \textbf{Perceived comprehension:}
    ``I understood the information presented in this explanation.''

    \item \textbf{Mental effort:}
    ``Understanding this explanation required a lot of mental effort.''

    \item \textbf{Perceived naturalness:}
    ``The sequence of information in this explanation felt natural.''
\end{enumerate}

Each item used a three-point agreement scale:
\[
1=\text{disagree}, \qquad
2=\text{neither agree nor disagree}, \qquad
3=\text{agree}.
\]

The three items were analyzed separately and treated as
single-item subjective indicators rather than as a composite
or validated psychometric scale.

\paragraph{Exposure duration and language versions.}

Explanation-display duration was treated as a complementary
descriptive measure rather than a direct measure of attention
or active reading. French and English formulations were
examined by the six pretest participants in each corresponding
language, and identified discrepancies were resolved before
the main evaluation. Complete bilingual questionnaire
materials, response options, and answer keys will be provided
in the supplementary material.

\subsection{Statistical Analysis}
\label{subsec:analysis-plan}

Question-level correctness on the objective-comprehension
questions was the primary outcome. Recommendation acceptance
and the three subjective judgments were secondary outcomes,
while exposure duration was treated descriptively.

\paragraph{Primary analysis.}

Question-level correctness was analyzed using a logistic
mixed-effects model with organization condition, question
identity, language, and trial position as fixed effects.
Trial position was treated as categorical, and question
identity accounted for differences in difficulty among the
eight objective questions. The initial model included random intercepts for participants
and for trials nested within participants. The trial-level
variance was estimated at the boundary ($<10^{-12}$),
resulting in a singular fit. The trial-level random intercept
was therefore removed. The final model retained a participant
random intercept and was non-singular. Three contrasts of primary interest compared OWN with DONOR,
POP, and PROFILE-LLM. Tests were two-sided and Holm-adjusted
\cite{Holm1979} across these three contrasts to control the
familywise error rate at 0.05. Results are reported as odds ratios, 95\% confidence intervals,
Holm-adjusted $p$-values, and estimated marginal probabilities
on the response scale, averaged over question identity, language,
and trial position. The descriptive
0--2 trial score did not replace the question-level analysis.

\paragraph{Question-type analysis.}

A secondary analysis tested whether condition effects differed
between factual and integration questions by adding a
condition-by-question-type interaction while retaining question
identity. A separate main effect of question type was not
interpreted because question type is determined by item identity. Differences between condition effects were evaluated through
the interaction rather than by comparing significance patterns
across question types. The interaction was evaluated using a
likelihood-ratio test.

\paragraph{Recommendation acceptance.}

The three-category acceptance outcome (``No,'' ``Uncertain,''
``Yes'') was analyzed using a multinomial mixed-effects
framework with organization condition, target situation,
language, and trial position as fixed effects and participant
as a random intercept. The overall effect of organization
condition was evaluated using a joint Wald test of the six
condition coefficients across the two non-reference outcome
categories.

\paragraph{Subjective outcomes.}

Perceived comprehension, mental effort, and perceived
naturalness were analyzed separately using cumulative-logit
mixed-effects models with organization condition, target
situation, language, and trial position as fixed effects and
participant as a random intercept. For each outcome, the
overall effect of organization condition was evaluated using
a likelihood-ratio test comparing models with and without the
condition term.

\paragraph{Multiplicity for secondary inference.}

One omnibus condition-by-question-type interaction test and
four omnibus condition tests for recommendation acceptance
and the three subjective outcomes formed a family of five
secondary tests. Holm's method~\cite{Holm1979} controlled the
familywise error rate across this family. Additional pairwise
contrasts or subgroup analyses were treated as exploratory.

\paragraph{Exposure duration.}

Exposure duration was summarized by condition using medians
and interquartile ranges. It was not included as a covariate
in the primary model because explanation organization could
itself affect duration; any additional duration analyses were
therefore treated as exploratory.

\paragraph{Incomplete responses and exclusions.}

Available valid objective responses were retained, and missing
responses were not recoded as incorrect. Exclusions were
limited to documented technical failures preventing reliable
delivery or response recording, records that could not be
linked to a valid participant and trial, and duplicate records. For duplicate submissions to the same question, the earliest
valid response was retained. Low correctness, uncertain
acceptance, and unusually short or long exposure duration were
not exclusion criteria, and objective questions were not treated
as attention checks. Missingness, if present, was described by
condition, language, and trial position. Because all 122 participants completed all four evaluation
trials, a separate participant-level complete-case analysis
did not define a different sample.

\paragraph{Sample size.}

The evaluation sample comprised the 122 participants with
complete historical profiles and was not determined by a
prospective power analysis. Inference therefore relies on
estimated effect sizes, 95\% confidence intervals, and the
mixed-effects analyses above rather than post-hoc observed
power.

\paragraph{Scope of inference.}

The analyses concern the complete explanation-organization
procedures rather than the independent causal contribution of
information selection, relative salience, or relational
organization. Comparisons involving PROFILE-LLM likewise
concern its complete procedure, including access to
target-context information, rather than an isolated
language-model effect. Differences between factual and integration performance may
inform interpretation but do not identify a unique mechanism
through which personalization operates. Conclusions are
therefore limited to the participants, situations, questions,
languages, and organization procedures represented in this
study.

\section{Results}
\label{sec:results}

All 122 participants completed the four evaluation trials, yielding
488 trials and 976 objective-comprehension responses. No responses
were missing, and no trial was excluded under the exclusion criteria
described above.

\noindent\textbf{RQ: Does organizing an explanation from interpretive
regularities derived from the user's prior interpretations improve
objective comprehension of the presented information compared with
alternative organization procedures?}

We report the primary objective-comprehension results first, followed
by question-type differences, secondary outcomes, exposure duration,
and exploratory consistency across contexts.

\subsection{Objective Comprehension}
\label{sec:results-objective}

Objective-comprehension accuracy was highest in the OWN condition.
Participants answered 211 of 244 questions correctly in OWN
(86.5\%), compared with 161 of 244 in DONOR (66.0\%),
156 of 244 in POP (63.9\%), and 178 of 244 in PROFILE-LLM
(73.0\%). The primary mixed-effects logistic model supported these comparisons
(Table~\ref{tab:primary-comprehension}). Compared with DONOR,
OWN was associated with 3.82 times higher odds of a correct response
(95\% CI [2.36, 6.18], $p_{\mathrm{Holm}}<0.001$).
The corresponding odds ratios were 4.23 relative to POP
(95\% CI [2.61, 6.84], $p_{\mathrm{Holm}}<0.001$)
and 2.67 relative to PROFILE-LLM
(95\% CI [1.64, 4.36], $p_{\mathrm{Holm}}<0.001$).

\begin{table}[t]
\centering
\small
\setlength{\tabcolsep}{4pt}
\caption{Primary objective-comprehension results. Odds ratios compare
OWN with each alternative organization procedure.}
\label{tab:primary-comprehension}
\begin{tabular}{lccc}
\toprule
Condition / Contrast & Accuracy & OR [95\% CI] & $p_{\mathrm{Holm}}$ \\
\midrule
OWN                  & 86.5\% & -- & -- \\
DONOR                & 66.0\% & -- & -- \\
POP                  & 63.9\% & -- & -- \\
PROFILE-LLM          & 73.0\% & -- & -- \\
\midrule
OWN vs. DONOR        & -- & 3.82 [2.36, 6.18] & $<0.001$ \\
OWN vs. POP          & -- & 4.23 [2.61, 6.84] & $<0.001$ \\
OWN vs. PROFILE-LLM  & -- & 2.67 [1.64, 4.36] & $<0.001$ \\
\bottomrule
\end{tabular}
\end{table}

\begin{figure}[t]
    \centering
    \includegraphics[width=\columnwidth]
    {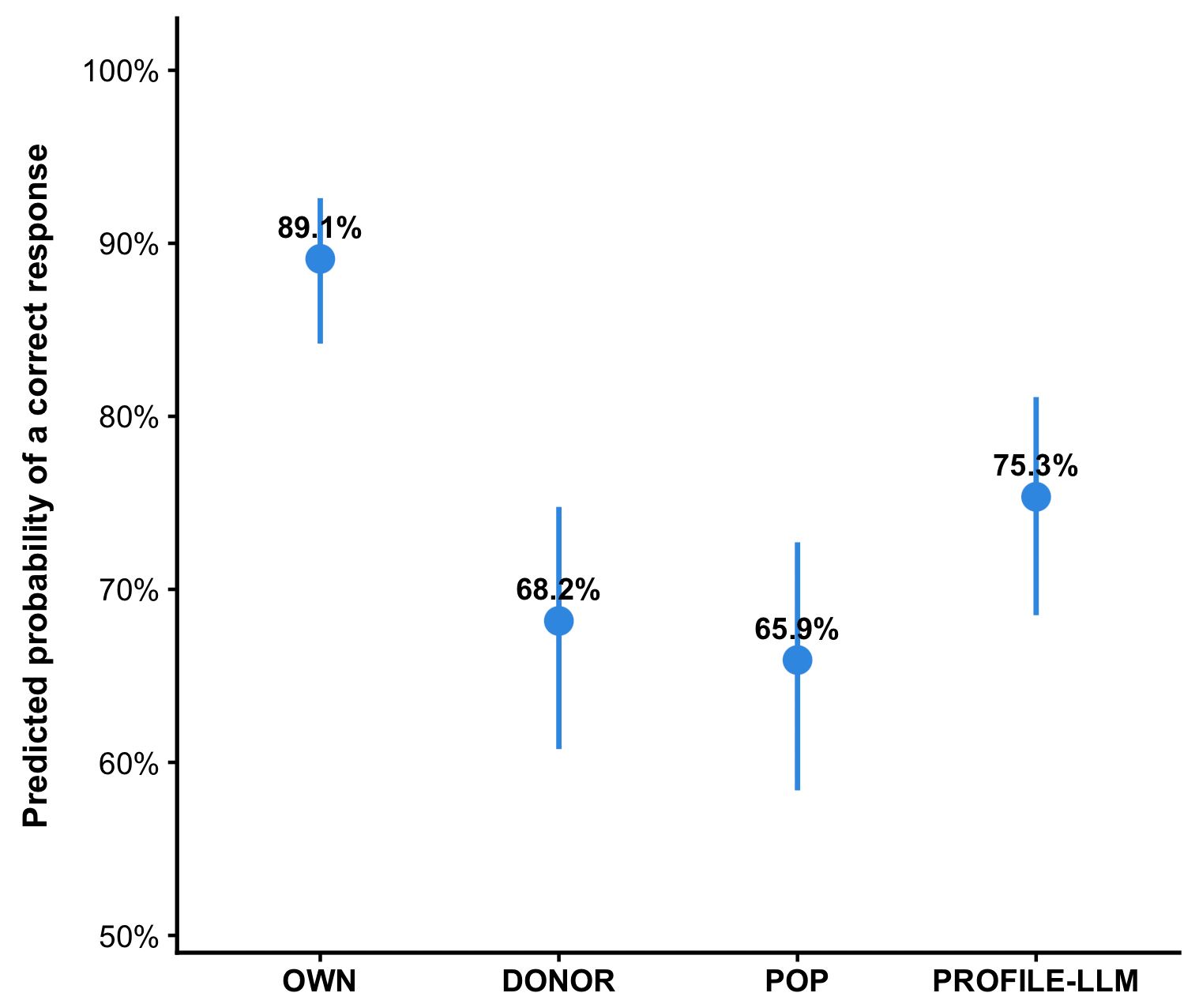}
  \caption{Estimated marginal probability of a correct
objective-comprehension response by organization condition from
the primary mixed-effects logistic model, averaged over question
identity, language, and trial position. Vertical lines indicate
95\% confidence intervals.}
    \Description{Predicted probabilities of a correct
    objective-comprehension response for the four organization
    conditions. OWN has the highest predicted probability,
    followed by PROFILE-LLM, DONOR, and POP. Vertical lines
    indicate 95 percent confidence intervals.}
    \label{fig:predicted-comprehension}
\end{figure}

Figure~\ref{fig:predicted-comprehension} shows the corresponding
model-predicted probabilities of a correct response with 95\%
confidence intervals. The predicted probability was highest for
OWN (89.1\%), followed by PROFILE-LLM (75.3\%), DONOR
(68.2\%), and POP (65.9\%).

Across all three primary contrasts, organizing the explanation
from the user's historical profile resulted in higher objective
comprehension of the presented information than each of the
alternative organization procedures tested.

\subsection{Factual and Integration Questions}
\label{sec:results-question-type}

OWN showed the highest descriptive accuracy for both question types.
For factual questions, accuracy was 89.3\% in OWN, compared with
72.1\% in DONOR, 68.9\% in POP, and 81.1\% in PROFILE-LLM.
For integration questions, accuracy was 83.6\% in OWN, compared with
59.8\% in DONOR, 59.0\% in POP, and 64.8\% in PROFILE-LLM
(Table~\ref{tab:question-type}).

\begin{table}[t]
\centering
\caption{Objective-comprehension accuracy by question type and
organization condition. Values are raw proportions.}
\label{tab:question-type}

\begin{tabular*}{\columnwidth}{
@{\extracolsep{\fill}}lcc@{}
}
\toprule
\textbf{Condition} &
\textbf{Factual} &
\textbf{Integration} \\
\midrule
OWN         & 89.3\% & 83.6\% \\
DONOR       & 72.1\% & 59.8\% \\
POP         & 68.9\% & 59.0\% \\
PROFILE-LLM & 81.1\% & 64.8\% \\
\bottomrule
\end{tabular*}

\end{table}

A likelihood-ratio test found no evidence that the effect of
organization condition differed by question type,
$\chi^{2}(3)=1.285$, $p=0.733$,
$p_{\mathrm{Holm}}=1.000$. Thus, although OWN had the highest
descriptive accuracy for both factual and integration questions,
the data did not indicate a differential condition effect across
the two question types.

\subsection{Subjective Judgments and Recommendation Acceptance}
\label{sec:results-secondary}

Subjective judgments differed descriptively across organization
conditions (Table~\ref{tab:secondary-outcomes}). For perceived
comprehension, 63.1\% of participants selected the highest response
category in OWN, compared with 32.8\% in DONOR, 34.4\% in POP,
and 44.3\% in PROFILE-LLM. A likelihood-ratio test indicated an
overall condition effect, $\chi^{2}(3)=29.27$, $p<0.001$,
$p_{\mathrm{Holm}}<0.001$. For perceived naturalness, the corresponding percentages were
54.9\% in OWN, 27.9\% in DONOR, 36.1\% in POP, and 37.7\% in
PROFILE-LLM. A likelihood-ratio test also indicated an overall
condition effect, $\chi^{2}(3)=29.29$, $p<0.001$,
$p_{\mathrm{Holm}}<0.001$. For mental effort, the highest response category was selected by
23.0\% of participants in OWN and 36.1\% in each of DONOR, POP,
and PROFILE-LLM. No reliable overall condition effect was
established after multiplicity correction
(likelihood-ratio test, $\chi^{2}(3)=7.41$, $p=0.060$,
$p_{\mathrm{Holm}}=0.180$).

\begin{table}[t]
\centering
\caption{Secondary outcomes by organization condition. Subjective
values report the percentage selecting response category 3;
acceptance reports the percentage selecting Yes.}
\label{tab:secondary-outcomes}

\begin{tabular*}{\columnwidth}{
@{\extracolsep{\fill}}lcccc@{}
}
\toprule
\textbf{Condition}
& \textbf{Compr.}
& \textbf{Effort}
& \textbf{Natural.}
& \textbf{Accept. Yes} \\
\midrule
OWN         & 63.1\% & 23.0\% & 54.9\% & 39.3\% \\
DONOR       & 32.8\% & 36.1\% & 27.9\% & 41.0\% \\
POP         & 34.4\% & 36.1\% & 36.1\% & 45.9\% \\
PROFILE-LLM & 44.3\% & 36.1\% & 37.7\% & 45.9\% \\
\bottomrule
\end{tabular*}

\smallskip
\footnotesize
\textit{Note.} Compr. = perceived comprehension;
Natural. = perceived naturalness;
Accept. Yes = percentage of Yes responses.
\end{table}

Recommendation acceptance showed a different descriptive pattern:
39.3\% of responses were Yes in OWN, 41.0\% in DONOR, 45.9\% in
POP, and 45.9\% in PROFILE-LLM. A joint Wald test found no evidence
of an overall condition effect on recommendation acceptance,
$\chi^{2}(6)=3.11$, $p=0.795$,
$p_{\mathrm{Holm}}=1.000$. Overall, OWN showed the highest descriptive percentages for perceived
comprehension and perceived naturalness, for which omnibus condition
effects were significant. No reliable overall condition effect was
established for mental effort after multiplicity correction or for
recommendation acceptance.

\subsection{Exposure Duration}
\label{sec:results-duration}

Median exposure duration ranged from 38.2 to 42.2~s across
conditions. It was 38.2~s in OWN (IQR [28.1, 50.3]),
39.8~s in DONOR (IQR [29.2, 51.7]), 42.2~s in POP
(IQR [30.8, 51.9]), and 40.6~s in PROFILE-LLM
(IQR [28.8, 53.8]). Exposure duration is reported descriptively only and was not
included as a primary covariate because it may itself be affected
by explanation organization.

\subsection{Exploratory Consistency Across Contexts}
\label{sec:results-consistency}

We descriptively examined whether the overall pattern observed in the
primary analysis was also visible across languages, target situations,
and trial positions (Table~\ref{tab:exploratory-consistency}). Across languages, OWN had the highest objective-comprehension accuracy
in both English (85.7\%) and French (86.9\%). The corresponding
PROFILE-LLM accuracies were 70.2\% and 74.4\%, respectively. Across target situations, OWN had the highest accuracy in T1
(91.7\%), T2 (89.7\%), and T3 (91.9\%). The pattern was not uniform:
in T4, PROFILE-LLM had the highest accuracy (78.3\%), compared with
73.4\% for OWN. Across trial positions, OWN had the highest accuracy in positions 1
(84.4\%), 2 (87.9\%), and 4 (91.9\%). At position 3,
PROFILE-LLM was slightly higher than OWN
(82.3\% vs.\ 81.7\%).

\begin{table}[t]
\centering
\caption{Exploratory objective-comprehension accuracy across
languages, target situations, and trial positions. Values are raw
proportions.}
\label{tab:exploratory-consistency}

\textbf{Language}

\begin{tabular*}{\columnwidth}{
@{\extracolsep{\fill}}lcc@{}
}
\toprule
\textbf{Condition} & \textbf{EN} & \textbf{FR} \\
\midrule
OWN         & 85.7\% & 86.9\% \\
DONOR       & 64.3\% & 66.9\% \\
POP         & 64.3\% & 63.7\% \\
PROFILE-LLM & 70.2\% & 74.4\% \\
\bottomrule
\end{tabular*}

\vspace{0.6em}

\textbf{Target situation}

\begin{tabular*}{\columnwidth}{
@{\extracolsep{\fill}}lcccc@{}
}
\toprule
\textbf{Condition} &
\textbf{T1} & \textbf{T2} & \textbf{T3} & \textbf{T4} \\
\midrule
OWN         & 91.7\% & 89.7\% & 91.9\% & 73.4\% \\
DONOR       & 65.6\% & 71.7\% & 67.2\% & 59.7\% \\
POP         & 62.9\% & 56.2\% & 75.0\% & 62.1\% \\
PROFILE-LLM & 74.1\% & 67.7\% & 71.9\% & 78.3\% \\
\bottomrule
\end{tabular*}

\vspace{0.6em}

\textbf{Trial position}

\begin{tabular*}{\columnwidth}{
@{\extracolsep{\fill}}lcccc@{}
}
\toprule
\textbf{Condition} &
\textbf{1} & \textbf{2} & \textbf{3} & \textbf{4} \\
\midrule
OWN         & 84.4\% & 87.9\% & 81.7\% & 91.9\% \\
DONOR       & 69.4\% & 57.8\% & 65.5\% & 71.7\% \\
POP         & 65.0\% & 59.7\% & 54.7\% & 77.6\% \\
PROFILE-LLM & 75.9\% & 65.0\% & 82.3\% & 68.8\% \\
\bottomrule
\end{tabular*}

\end{table}

Overall, the exploratory analyses were broadly consistent with the
primary pattern while also revealing context-specific exceptions.

\section{Discussion}
\label{sec:discussion}

\paragraph{Historical profiles as an organizing signal.}

Explanations organized from the user's own historical profile yielded
higher objective comprehension than the three alternative organization
procedures. Because explanation content and wording were held constant,
the observed advantage concerns information organization rather than
the provision of additional explanatory content. The findings therefore
support the use of historical profiles derived from prior interpretations
as an actionable signal for organizing subsequent explanations.

\paragraph{What the procedure comparisons establish.}

The comparisons with DONOR and POP indicate an advantage for using the
user's own historical profile rather than another participant's profile
or a leave-one-participant-out population profile. OWN also outperformed
the target-aware PROFILE-LLM procedure tested here, although both used
the user's historical profile. This should not be interpreted as a
general superiority of deterministic rules over language models, since
the two are complete organization procedures that use profile and target
information differently. Likewise, the design does not isolate the
independent contributions of information selection, relative salience,
and relational organization. Because historical salience accounted for
most of the observed sentence-order structure, the comprehension
advantage should not be interpreted as evidence that the three profile
components contributed equally.

\paragraph{Objective comprehension, subjective experience, and acceptance.}

Perceived comprehension and perceived naturalness showed reliable
condition effects, whereas no reliable overall condition effect was
established for mental effort after multiplicity correction or for
recommendation acceptance. The absence of a condition-by-question-type
interaction also provided no evidence that the comprehension advantage
differed between factual and integration questions. These secondary
outcomes delimit the primary result and do not identify a specific
cognitive mechanism.

\paragraph{Contextual variation.}

The OWN advantage was not uniform across all exploratory subsets:
PROFILE-LLM was descriptively higher in T4 and slightly higher at trial
position 3. Because the study was not designed to identify the causes of
these reversals, a post-hoc mechanistic explanation would be unwarranted.
These exceptions instead motivate examining when historical-profile-based
organization is most beneficial and when target-adaptive procedures may
be preferable.

\paragraph{Relation to prior work.}

Prior work has personalized explanatory content, presentation, and
organization \cite{SchneiderHandali2019,KoukiEtAl2019}. Explanation
ordering under time constraints has also been guided by explicitly
elicited user importance
\cite{MundukuNegre2026StructureMatters}. The present study differs by
deriving the organizing signal from prior open-ended interpretations and
by evaluating objective comprehension rather than recommendation
acceptance as the primary outcome.

\paragraph{Implications.}

The findings suggest that user representations may capture not only what
information a person prefers or selects, but also regularities in how
information is emphasized and organized across prior interpretations.
Such representations can inform presentation while leaving the underlying
content unchanged. Explanation organization may therefore serve as a
personalization dimension under the controlled procedures studied here,
although its generalization to richer adaptive interfaces and longer-term
use remains to be established.

\subsection{Limitations}
\label{sec:limitations}

Several limitations delimit the scope of these findings. Participants
were recruited through convenience channels, and the study was limited
to French and English, four controlled target situations, six predefined
functional roles, and fixed sentence-level explanations. Of the 135
participants who began profile construction, 122 provided the four
interpretations required for evaluation, so the analytical sample also
reflects completion of the profile-construction phase. Each historical profile was derived from only four prior interpretations,
and the automatic extraction procedure, although fixed before evaluation,
was not independently revalidated on the present corpus. OWN should also
be interpreted as a complete explanation-organization procedure: the
design does not isolate the independent contributions of information
selection, relative salience, and relational organization. Moreover,
historical salience accounted for most of the observed sentence-order
structure, while the relational grouping step modified only a subset of
orders. The observed advantage therefore should not be interpreted as
evidence that all profile components contributed equally. Relational organization was only partially operationalized through
co-mobilization. This indicator is symmetric and untyped and does not
identify semantic, causal, directional, or otherwise typed relations.
A positive historical co-mobilization coordinate may also be supported
by only one source episode. PROFILE-LLM represents one specific model and prompting procedure and
also conditions its ordering on target-specific information. Its
comparison with OWN therefore does not establish a general advantage
over language-model-based personalization. The study also did not include
a direct preference-based personalization condition; therefore, it does
not establish whether historical profiles provide benefits beyond
personalization based on explicitly elicited preferences. Finally, objective comprehension was assessed through eight
multiple-choice questions after the explanation had been removed, so
performance reflects both comprehension and short-term retention and
does not identify a unique cognitive mechanism. The subjective outcomes
were single-item, three-point task-specific indicators rather than
validated psychometric scales. Broader and longer-term studies are
needed to assess generalization to less controlled information-interaction
settings.

\section{Conclusion}
\label{sec:conclusion}

This study examined whether historical profiles derived from prior
interpretations can guide the organization of personalized explanations.
Explanations organized from the user's own historical profile yielded
higher post-exposure objective comprehension than DONOR, POP, and
PROFILE-LLM while explanation content and wording remained fixed.
Because comprehension was assessed after the explanation had been
removed, performance also reflects short-term retention. These findings indicate that historical profiles can provide an
actionable signal for explanation organization. Higher objective
comprehension did not coincide with a detectable overall condition
effect on recommendation acceptance. More broadly, the results suggest
that personalization can operate not only through content selection but
also through information organization. Future work should examine this
approach in broader populations, richer explanation settings, and
longer-term deployments.

\bibliographystyle{ACM-Reference-Format}
\bibliography{references_paper2_clean_complete}

\end{document}